# Clinicians' Interpretation and Preferences for Survival Data Visualisation: A Pre–Post Study Comparing Kaplan–Meier and Mean Residual Life Plots

Victor Pacifique Rwandarwacu[1,*] \ [1]School of Mathematical and Statistical Sciences, \ University of Galway, Ireland \ *Corresponding author: `v.rwandarwacu1@universityofgalway.ie`

2025-11-11

**Abstract**

Effective visualisation of survival data is key for both clinician interpretation and patient communication. While a Kaplan Meier (KM) is a popular and standard plot, Mean Residual Life (MRL) plots may offer a more informed visualisation of prognosis on time scale. However, little is known about clinicians' knowledge and preferences regarding these alternative visualisations. In this pre–post (Pilot) cross-sectional survey, 32 participants including medical students and doctors voluntarily completed an online questionnaire where they were asked to interpret four survival visualisation types (KM plot, survival difference, MRL plot, and difference in MRL) before and after a brief learning section. Outcomes assessed included interpretation accuracy, learning gains, and plot preferences for clinician's understanding and use with patients.

Overall, the pre-learning (baseline) interpretation accuracy was 50% [37.5-62.5], and it was higher for the Kaplan-Meier plot than the Mean Residual Life (MRL) plot at 56.2% [37.5-71.9] and 43.8% [28.1-62.5], respectively. Following the learning section, the overall accuracy improved significantly to 81.2% [68.8 - 92.2] (p = 0.002). The most significant improvement was observed for the MRL plot (+37.5 percentage points, p = 0.010), which achieved a post-learning accuracy of 81.2%, similar to that of KM. Kaplan–Meier plots were the most preferred for both ease of understanding by clinicians (59%) and patient communication (53%). There was a noticeable improvement in interpretation among participants with lower self-rated knowledge, especially for MRL visualisation. Although KM is popular in medical literature, MRL plots can be rapidly understood by clinicians with minimal instruction. These findings echo the integration of MRL visualisations into clinical dashboards and medical education to enhance survival data communication. There is a need for large-scale study in a controlled setting to validate these findings.

**Availability**: Code available at https://github.com/rwandarwacu1/Msc_thesis_survival

# 1 Introduction

## 1.1 Background

Survival analysis, also referred to as time-to-event data analysis, is the analysis of data in the form of times from a well-defined time origin until the occurrence of some particular event or end point. Survival analysis is a critical branch of statistics that specifically deals with the modeling of time until an event occurs. These events may range from biological outcomes such as death, remission, or relapse, to engineering failures, time to default in finance, or germination in agriculture (Peace [2009]). The distinct feature of survival data is censoring, wherein the event of interest has not been observed for all subjects during the study period. This demands specialized statistical methods to handle incomplete observations and skewed distributions (Collett [2023]; Bradburn et al. [2003]).

In medical research, survival analysis is used extensively to study disease prognosis, treatment effectiveness, and risk prediction. The Kaplan–Meier (KM) estimator remains one of the most commonly used tools for visualising survival functions due to its simplicity and non-parametric formulation (Klein and Moeschberger [2006]). However, despite widespread usage, KM plots can be difficult to interpret for individuals without statistical background, particularly in clinical practice and patient-centered communication contexts (Hawkins et al. [2005]; Trevena et al. [2013]).

### 1.1.1 Survival Function

The survival function $S(t)$ is the probability that the survival time is greater than or equal to time $t$, which is the observed value of random variable $T$ with distribution function $F(t)$ [Collett, 2023]. It is a non-increasing function ranging between 1 and 0.

$$S(t) = \mathrm{P}(T \geqslant t) = 1 - F(t) \tag{1}$$

The cumulative distribution function is the complement of the survival function; it gives the probability that the event has occurred by time $t$.

$$F(t) = \mathrm{P}(T < t) = \int_0^t f(u)\,\mathrm{d}u \tag{2}$$

The Kaplan-Meier (KM) estimator is a non-parametric method used to estimate the survival function from observed survival times, especially when data are censored. The resulting estimate is a stepwise approximation of the true survival function on the probability scale, graphically represented further in the next section.

The KM estimate of the survival function at the $k$-th interval is given by:

$$\hat{S}(t) = \prod_{j=1}^{k}\left(\frac{n_j - d_j}{n_j}\right) \tag{3}$$

For $t_{(k)} \leqslant t < t_{(k+1)}$, $k = 1, 2, ..., r$, with $\hat{S}(t) = 1$ for $t < t_{(1)}$, where $t_{r+1}$ is taken to be $\infty$. Here, $d_j$ denotes the number of deaths in this interval, $n_j$ is the number of individuals alive just before $t_{(j)}$, and $d_j$ events occur at $t_{(j)}$ [Collett, 2023].

The key analytical problem in survival analysis is censoring. The most common censoring mechanisms are right and left censoring; however, we will focus on right censoring, which occurs when a subject has been followed up, but the exact survival time is known to be greater than the last follow-up time. For example, in a study in which leukemia patients are followed until they go out of remission, if the study ends while the patient is still in remission, then the patient's survival time is considered censored (Administrative censoring). Moreover, censoring may also occur when a person is lost to follow-up during the study period or withdrawn from the study due to death or other reasons (Kleinbaum and Klein [2012]).

### 1.1.2 Mean Residual Life (MRL) Function

The MRL function estimates how much longer, on average, a subject is expected to survive given that the subject has survived until time $t$.

Mathematically, the MRL function is given by :

$$m(t) = E(T - t \mid T > t) = \frac{1}{S(t)} \int_t^\infty S(s)\, ds \qquad (4)$$

where $T$ is survival time and $S(t)$ the survival function.

Formula 4 demonstrates that the MRL is the area under the survivor function beyond time $t$, normalized by the conditional probability of surviving beyond that point.There are various estimators, but most of them fail for right censored data. Unlike the Kaplan–Meier estimator for survival probabilities, MRL estimation requires a complete understanding of the upper tail of the survival distribution. Because censoring obscures this region, reliable estimation demands additional assumptions or supplementary modelling,thus motivating the development of hybrid and semi-parametric approaches.

Previous studies have suggested different hybrid approaches, especially for parametric options in the upper tail, such as the exponential method (Su et al. [2012]). However, these approaches are insufficiently flexible when the tail behavior is unknown a priori. Notably, authors including Gelber, Goldhirsch, and Cole (Gelber et al. [1993]), as well as Gong and Fang (Gong and Fang [2012]), have proposed the use of goodness-of-fit procedures to guide selection of suitable parametric models for the tail estimation. However, the selection of candidate distributions is subject, and the large variability associated with the Kaplan–Meier curve could compromise the choice of the appropriate model.

To address these limitations, innovative statistical methodologies have emerged. A hybrid semi-parametric approach proposed by Alvarez-Iglesias et al. (2015) leverages both non-parametric estimation for the observed data and parametric modeling for the unobserved(censored) tail of the survival distribution. This method conceptually divides the MRL estimation task into two distinct components:

- **Below a threshold**: The MRL is estimated non-parametrically using the Kaplan–Meier estimator. This part captures the behavior of the data that is well-supported by observed events.
- **Above the threshold**: The tail of the survival function is modeled using extreme value theory (EVT), specifically by fitting a Generalized Pareto Distribution (GPD) to the excess times.

This combination addresses the issue of right-censoring by extrapolating the survival distribution beyond the last observed time point, using a model justified by the Extreme Value Theory (EVT).

The GPD is selected for its flexibility—it encompasses a range of tail behaviors from light (e.g.,exponential) to heavy-tailed distributions. Its defining parameters, the scale $\sigma$ and shape $\xi$, are estimated via maximum likelihood using the observed data above the threshold $u$. Importantly, EVT provides theoretical justification for using the GPD in this setting, ensuring that the extrapolation is statistically valid under broad conditions.

The threshold plays a critical role. It should be high enough to ensure that the EVT approximation holds, but low enough to retain sufficient data for stable estimation. Diagnostic tools or quantile-based heuristics (e.g., the 80th percentile of observed uncensored times) are frequently employed to guide this selection.

**1.1.2.1 Description of hybrid semi-parametric method** Let $T$ be the true survival time and let $C$ be the censoring time, and assume that $T$ and $C$ are independent.

For an individual $i$, the observed pair is $(X_i, \delta_i)$, where $X_i = \min(T_i, C_i)$ and $\delta_i = I_{T_i < C_i}$.

Let $\{(x_1, \delta_1), (x_2, \delta_2), \ldots, (x_n, \delta_n)\}$ be a random sample of observed times and censoring indicators.

Let $T^*$ be the maximum observed time. The estimate of the Mean Residual Life (MRL) function at time $t$, denoted $\hat{m}(t)$, is obtained as follows:

1. Choose a threshold $u$ and select the observations $\xi$ that are between $u$ and $T^*$. Let $m$ denote the number of these observations.

2. Calculate the maximum likelihood estimates of the parameters $\xi$ and $\sigma_u$ of the Generalized Pareto Distribution (GPD) using the $m$ observations that are between $u$ and $T^*$. The censored likelihood function is

$$L(\xi, \sigma_u; t_i, \delta_i) = \prod_{i=1}^{m} \left( \frac{1}{\sigma_u} \left[ 1 + \frac{\xi t_i}{\sigma_u} \right]^{-1/\xi} \right)^{\delta_i} \left( \left[ 1 + \frac{\xi t_i}{\sigma_u} \right]^{-1/\xi} \right)^{1-\delta_i} \quad (5)$$

where $t_i$ are the excesses above the threshold $u$, that is, $t_i = x_i - u$. This function can be maximized using standard optimization methods.

3. Estimate the restricted MRL below $u$ using the Kaplan-Meier estimator. This estimate will be called $\hat{m}_{\text{KM}}(t)$.
4. Estimate MRL at time $u$ for the GPD:

$$\hat{m}(u) = \frac{\hat{\sigma}_u}{1 - \hat{\xi}} \quad (6)$$

where $\hat{\xi}$ and $\hat{\sigma}_u$ are the maximum likelihood estimates from step 2.

5. Finally, the full MRL estimate at any time $t$ is obtained by summing the estimated area under the Kaplan–Meier curve from $t$ to $u$, and the area under the extrapolated tail from $u$ to infinity, weighted by survival probabilities:

$$\hat{m}(t) = \hat{m}_{\text{KM}}(t) + \hat{m}(u) \cdot \frac{\hat{S}_{\text{KM}}(u)}{\hat{S}_{\text{KM}}(t)} \quad (7)$$

This method allows for a smooth and interpretable estimation of the MRL, even when censoring prevents direct estimation using classical techniques.

## 1.2 Example dataset

The dataset used to generate figure 1 to 4 was drawn from the Cancer Genome Atlas – Breast Cancer (TCGA-BRCA) database, a well-curated, open access repository of longitudinal survival and clinical data (Can).

To simulate realistic clinical subgroup comparisons, subjects were stratified based on binary pathological stages. The plot types were constructed using R and later used in the survey for interprtation and ranking by clinicians which is further discussed in Methodology section. Generated plots include :

- **Kaplan-Meier (KM) plot**: Stepwise curve showing estimated survival probability over time
- **Survival difference plot**: Curve showing pointwise difference in survival probabilities between groups. The survival difference between two groups $A$ and $B$ at time $t$ is given by:

$$\Delta S(t) = S_A(t) - S_B(t) \quad (8)$$

Where: $S_A(t)$ is the survival function for group $A$, $S_B(t)$ is the survival function for group $B$.

- **Mean Residual Life (MRL) plot**: Mean residual life at time t; expected remaining lifespan given survival to t

- **Difference in Mean Residual Life (MRL) plot** : Difference in MRL estimates between two groups over time

MRL plots were estimated using a hybrid semi parametric method proposed by Alvarez-Iglesias et al. [2015]

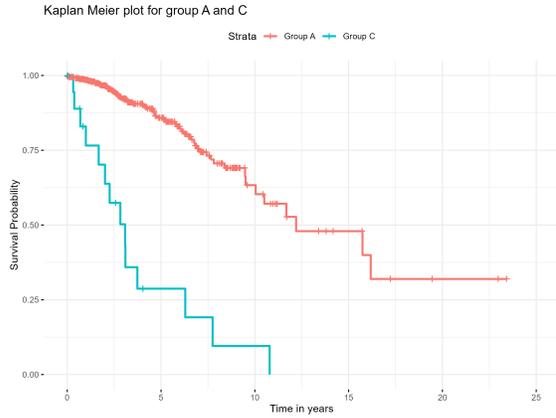

Figure 1: Kaplan Meier plot

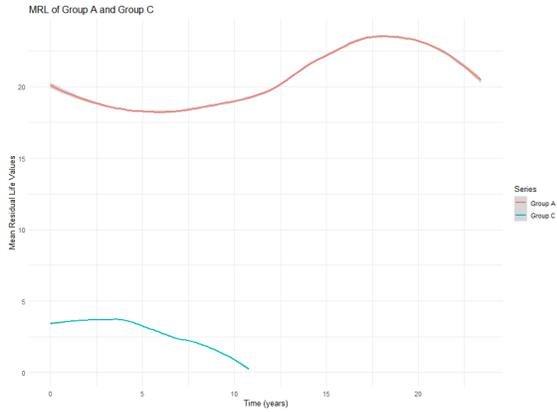

Figure 2: Mean Residual Life plot

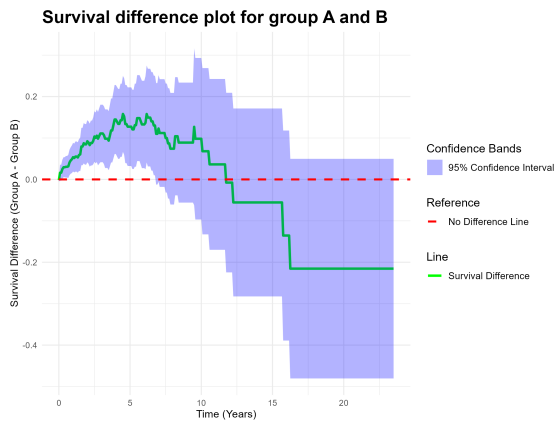

Figure 3: Difference in survival plot

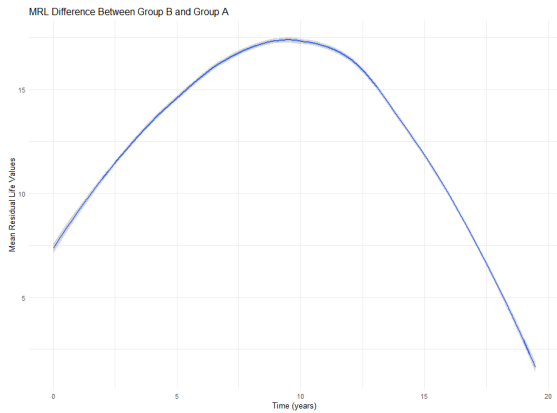

Figure 4: Difference in MRL plot

## 1.3 Survival Visualisation

While KM and survival difference plots are informative, they operate on a probability scale, which may not resonate intuitively with human perceptions of remaining lifespan or life expectancy (Spiegelhalter et al. [2011]). This limitation is especially relevant when clinicians need to translate survival statistics into terms understandable by patients, such as "How much longer can I expect to live?" (Pencina et al. [2009]).

The Mean Residual Life (MRL) function is a valuable alternative in such scenarios. Unlike survival probabilities or hazard rates, MRL explicitly quantifies prognosis in time units, which may better support clinical interpretation and patient discussions (Wang et al. [2017]; Wang et al. [2023]). The hybrid estimator discussed herein presents a flexible yet theoretically grounded framework for estimating MRL in the presence of censoring and has been implemented in practical tools such as interactive Shiny web applications designed for translational statistics (Alvarez-Iglesias; Jalali [2018]).

For the graphical comparison of time-to-event outcomes across different groups, survival difference, and survival ratio plots are the most popular alternatives. The survival ratio plot, in particular, facilitates the identification of temporal windows during which survival outcomes diverge between distinct populations or interventions. Using resampling techniques, permutation envelopes are generated as reference bands to investigate whether the observed differences reflect true effects or arise from random variation is due to sampling variation or a possible real effect (Newell et al. [2006]). Nonetheless, these visualisations continue to express outcomes on the probability scale.

The survival ratio between two groups $A$ and $B$ at time $t$ is given by:

$$\text{Survival Ratio}(t) = \frac{S_A(t)}{S_B(t)} \tag{9}$$

where: $S_A(t)$ is the survival function for group $A$, $S_B(t)$ is the survival function for group $B$.

The consensus study on visualising harm in clinical trials suggested the top most recommended plot across diverse data categories, including binary, etc., and specifically for time to event data, Kaplan Meier, survival ratio plot, and mean cumulative plots were the only recommended options (Phillips et al. [2022]) . The mean residual life plot did not figure among the top choices.

Sengupta (Sengupta [1995]) has emphasized the importance of improved graphical representations for censored survival data, advocating for more intuitive visual tools that can bridge the gap between quantitative accuracy and clinical usability.

Notably, the relevance of MRL visualizations has been explored in various health-related domains, including oncology, transplant medicine, and infectious disease prognosis (Achilonu [2017]; Qayoom et al. [2025]). However, it is rarely used in clinical settings (Alvarez-Iglesias et al. [2015]), and the actual interpretation and preference patterns among clinicians—particularly when exposed to both traditional and MRL-based plots—remain under-explored.

### 1.4 Clinician Interpretation and Communication Gaps

As the healthcare sector increasingly embrace digital transformation data-driven decision-making, there is a pressing need to ensure that the output of sophisticated statistical models is both interpretable and actionable by frontline clinicians (Bélanger et al. [2011]), which aligns with the translational statistics concept, which encourages the dissemination of statistical findings in a manner that clinicians with a limited statistical background can easily grasp and apply (McCabe and Newell [2022]). Despite substantial advances in modeling, the usability of survival plots in real-world clinical communication remains limited (Cox et al. [2009]). Poor comprehension of survival data can hinder patient counseling, risk stratification, and even shared decision-making (Trevena et al. [2013]).

There is increasing acknowledgement that MRL plots, by shifting the focus from probabilities to remaining life expectancy, could be a better communication tool. Yet, formal assessment of their interpretability, preference, and perceived usefulness among medical practitioners has not been rigorously conducted in the literature.

### 1.5 Objectives of the study

This study aims to assess the baseline knowledge of medical students and doctors on survival plot interpretation and how their comprehension evolves post-learning. It also compares MRL visualisations with conventional survival analysis plots, specifically Kaplan Meier curves, in terms of interpretation accuracy and clinician preferences for communicating prognosis to patients.

# 2 Methods

## 2.1 study design overview

This is a mixed methods study. This pilot study adopted a pre–post learning cross-sectional survey design to assess how medical doctors and students interpret and evaluate various visualisations of time-to-event data including Kaplan–Meier (KM), Mean Residual Life (MRL) plots. The survey was deployed using an online Google Form, accessible via both mobile and desktop devices, and was open for responses over a period of six weeks.

## 2.2 Participants and Recruitment

Eligible participants included medical students and doctors. The study imposed no restrictions based on geographical location or medical specialty. The recruitment was carried through convenience sampling, using professional groups Social media platforms targeting health professionals.

Participation was entirely voluntary, and no incentives were offered. A total of 32 complete responses were included in the final analysis after excluding incomplete submissions. Ethical approval was not required for this pilot study.

## 2.3 Survey structure

The structured questionnaire had five sections :

- **(1)** Demographic and self rated score survival knowledge: Participants were asked to report their country of medical training,level of clinical training and self-assessed familiarity with survival analysis (Likert scale 1 to 10).

- **(2)** Pre learning (baseline) plot interpretation: Participants were shown four distinct survival visualisations and asked to provide brief written interpretations. This section aimed to assess their baseline understanding.

- **(3)** Learning section: The correct interpretations were provided for the plots given in section 2.

- **(4)** Post-learning plot(test) interpretation: Respondents were given a new set of plots, similar in format but based on different data. They were asked to answer the same interpretation questions as before, enabling comparison of pre- and post-learning accuracy.

- **(5)** Ranking visualisations:Participants ranked the plots for easy interpretability by clinicians (themselves) and use for patient communication in consultation.

## 2.4 Statistical Analysis

Each interpretation response was scored as correct or incorrect basing on respondent's ability to identify visually a group with better survival. Accuracy score was calculated per plot and per participant for both the pre- and post-learning phases, focusing specifically on Kaplan Meier and MRL plot. To assess the statistical significance of changes in accuracy post learning section, Paired McNemar's tests were applied. Descriptive statistics were used to summarise participant demographics, self-reported knowledge levels, and ranking outcomes. Learning gain was calculated as the percent increase in correct interpretation per participant.

Exploratory regression tree analysis was used to identify predictors associated with greater improvement among pre-learning score, qualification, and self rated knowledge. Moreover, classification tree was fitted to examine variables associated with accurate interpretation of MRL plot post learning, representing the novel visualisation under evaluation.

Table 1: Respondents' characteristics

|  | Overall |
|---|---|
|  | (N=32) |
| **Country** |  |
|   Ethiopia | 2 (6.3%) |
|   Germany | 1 (3.1%) |
|   Ireland | 3 (9.4%) |
|   Italy | 1 (3.1%) |
|   Portugal | 1 (3.1%) |
|   Rwanda | 20 (62.5%) |
|   Spain | 1 (3.1%) |
|   Switzerland | 3 (9.4%) |
| **Self_reported_Knowledge** |  |
|   Mean (SD) | 4.09 (2.37) |
|   Median [Min, Max] | 4.00 [1.00, 9.00] |
| **Qualification** |  |
|   Medical Doctor (General Practitioner) | 14 (43.8%) |
|   Medical Doctor (Specialist or postgraduate) | 15 (46.9%) |
|   Medical student | 3 (9.4%) |

NVivo was used to code correct and incorrect interpretations and then R for complete data analysis, visualizations by ggplot2 library, statistical tests and ranking summaries.

## 2.5 Outcome measures

This study assessed interpretation accuracy, learning gain, and Plot ranking for clinician ease of understanding and patient education. It also explored the predictors of improvement in the post-learning section.

# 3 Results

## 3.1 Participants characteristics

A total of 32 participants from nine countries completed the survey. The most common country of medical training was Rwanda (62.5%), followed by Ireland (9.4%) and Switzerland (9.4%). The majority of participants were either medical doctors with specialty training or residents (46.9%) or general practitioners (43.8%) (Table 1).

Participants self-rated their prior knowledge of survival analysis with a mean of 4.09 (SD = 2.37) on a 10-point Likert scale, with a median of 4. As shown in Figure 5, 75% of respondents rated their knowledge five or less, indicating generally limited prior familiarity with survival methods.

## 3.2 Overall Interpretation Accuracy

At baseline, the overall pre-learning interpretation accuracy for KM and MRL plots was 50.0% (95% CI [37.5 - 62.5]). Figure 6 illustrates the relationship between self-rated knowledge of survival analysis (mean =

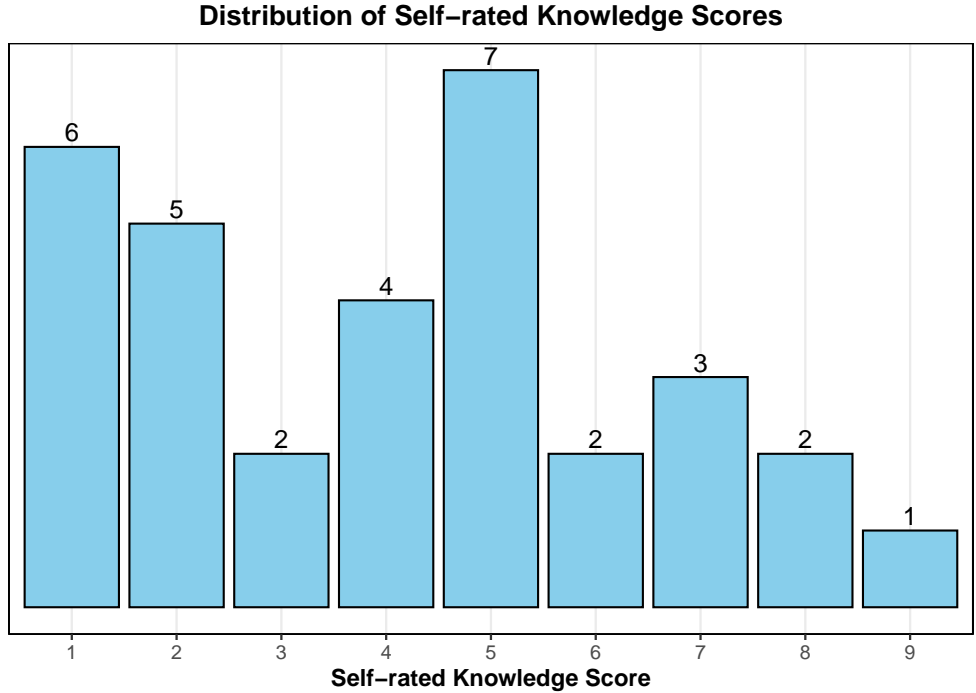

Figure 5: Distribution of self-rated knowledge scores for survival analysis: 5 and 1 are the most frequent score, 7 and 6 respectively.

4.09) and pre-learning accuracy. Participants with higher self-rated knowledge (above five) have comparable performance to those with lower self-rated knowledge (below five). However, those with the lowest self-rated knowledge, 1/10, have the lowest score, 25%.

Following the learning section, post-learning accuracy improved significantly to 81.2% (95% CI [68.8 - 92.2]). This represents a mean learning gain of +31.2 percentage points (95% CI [15.6 - 46.9]), which was statistically significant (Wilcoxon signed-rank test, p = 0.002), as shown in Figure 7 and Table 2. These results indicate that the brief learning exercise was effective in improving participants' ability to interpret survival visualisations.

## 3.3 Improvement in Interpretation accuracy per Plot Type

Interpretation improvements varied across both KM and MRL plots, as demonstrated in Figure 8, and Figure 9, and summarised in Table 3. Kaplan–Meier plots, which were the most familiar to participants, showed an improvement in accuracy from 56.2% pre-learning to 81.2% post-learning, equivalent to a gain of 25.0 percentage points (p = 0.027). The novel MRL plot showed an improvement, with accuracy increasing from 43.8% to 81.2%, representing a substantial gain of 37.5 percentage points (p = 0.010).

An alluvial diagram (Figure 9) illustrates the transitions in interpretation accuracy for both plot types. The most shifts from incorrect to correct interpretations were observed for the MRL plot, reflecting the impact of the learning section on an individual level and indicating a positive learning effect.

## 3.4 Learning Gain Among Participants with Lower Self-Rated Knowledge

The effect of the learning section was more pronounced among participants with lower self-rated knowledge of survival analysis. Figure 10 shows that participants who rated their prior knowledge below 5 achieved significant improvement, especially in interpreting MRL plots. This finding highlights that even clinicians and students with limited statistical knowledge can rapidly improve their understanding of advanced survival visualisations when provided with well-suited materials and contexts.

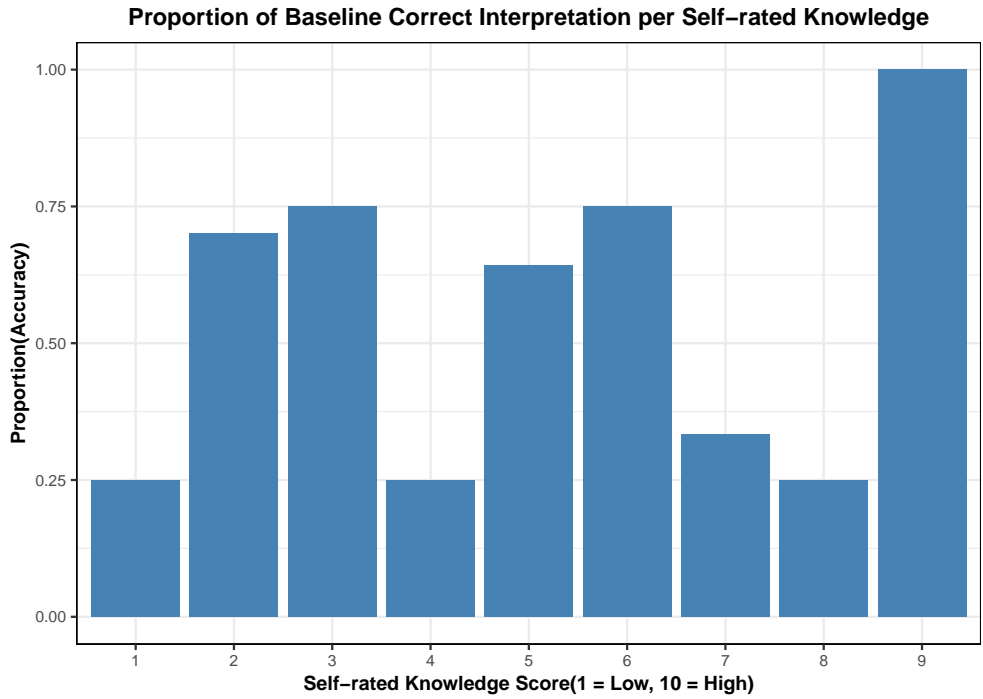

Figure 6: Proportion of Correct Interpretation per Self-rated Knowledge

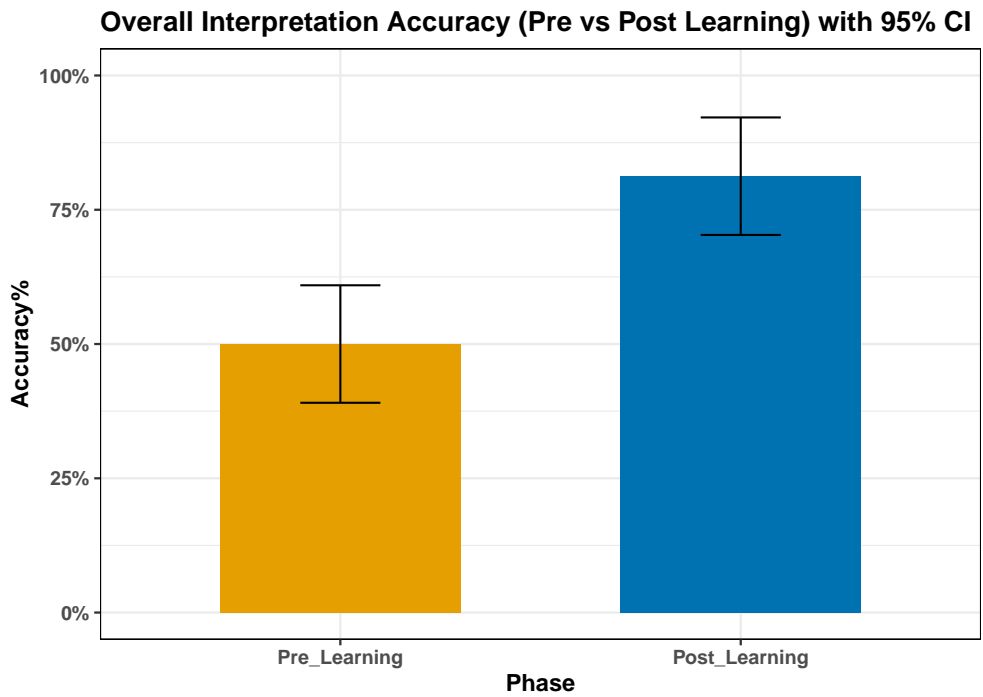

Figure 7: Overall Proportion of Correct Interpretations (with 95% CI),

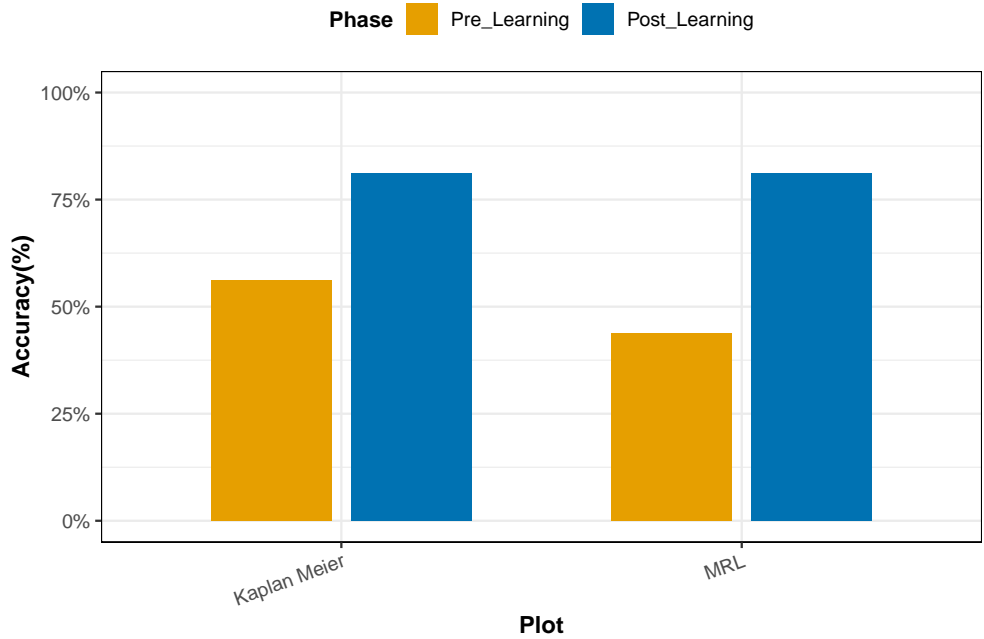

Figure 8: Proportion of Correct Interpretations (Pre vs Post Learning) by plot

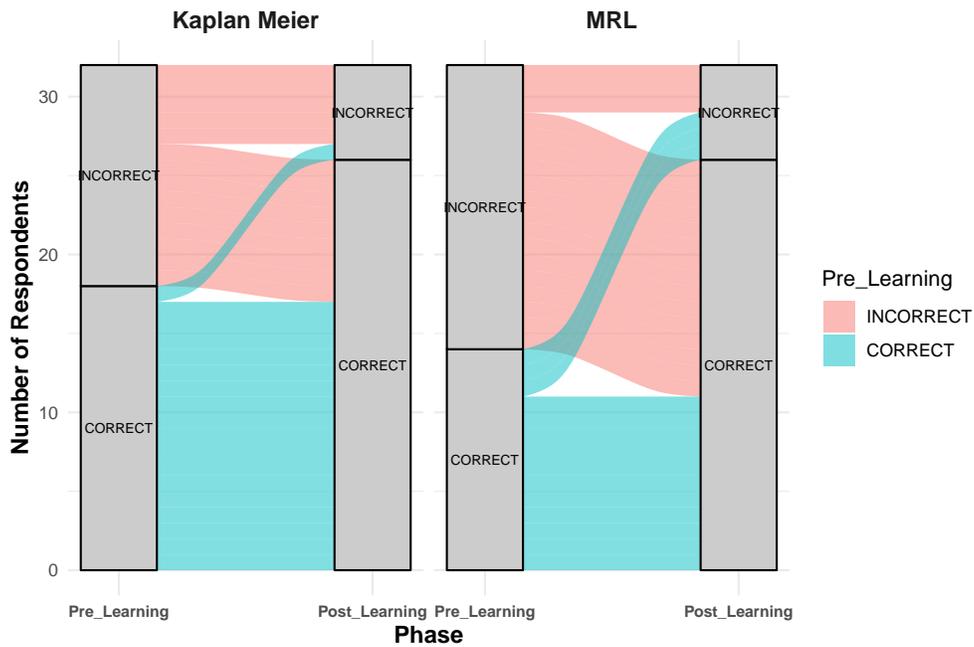

Figure 9: Alluvial Diagram of Changes in Interpretation per Plot Type

Table 2: Overall Summary of Key Outcomes (with 95% CIs, Wilcoxon p-value)

| Metric | Overall |
|---|---|
| Pre-learning accuracy | 50.0% [37.5 - 62.5] |
| Post-learning accuracy | 81.2% [68.8 - 92.2] |
| Learning gain (points) | +31.2 [14.1 - 46.9] |
| Learning gain p-value (Wilcoxon) | p = 0.002 |

Table 3: Summary of Key Outcomes (with 95% CIs) Per Plot

| Metric | KM Plot | MRL Plot |
|---|---|---|
| Pre-learning accuracy | 56.2% [40.6 - 71.9] | 43.8% [28.1 - 59.4] |
| Post-learning accuracy | 81.2% [65.6 - 93.8] | 81.2% [68.8 - 93.8] |
| Learning gain (points) | +25.0 [6.2 - 43.8] | +37.5 [15.6 - 59.4] |
| Learning gain p-value | p = 0.027 | p = 0.01 |
| Top-ranked for ease to communicate to patients | 53% | 9% |
| Top-ranked for ease of understanding by clinicians | 59% | 6% |

## 3.5 Exploratory Analysis of predictors of Accuracy

Regression tree analysis (Figure 11) identified the pre-learning total correct score and participant qualification, such as being a medical specialist or student, as key predictors of overall learning gain among the respondents. In general, participants with lower baseline scores demonstrated greater improvement following the learning section.

Figure 12 further illustrates that a combination of higher self-rated knowledge and qualification status best predicted post-learning accuracy for the novel MRL plot. Participants with high self-rated scores and general practitioners were more likely to achieve higher interpretation accuracy for MRL plots in the post-learning section, underscoring the accessibility of MRL visualisations when supported by structured learning in both clinical and academic settings.

## 3.6 Clinicians ranking preferences

Participants were also asked to rank four visualisation types based on their perceived usefulness for clinical interpretation and for communicating the clinical outcome to patients. Figure 13 displays these preferences using a heat map. For clinical use, the Kaplan–Meier plot was the most preferred, selected as the top choice by 59% of respondents (Table 3). The MRL plot was only the top choice for 6% of respondents. Regarding patient communication, KM was still the most preferred option, with 53% of participants selecting it as the most helpful format for explaining clinical outcomes to patients. The preference for MRL plots has also increased slightly to 9% (Table 3). The interactive shiny dashboard is accessible through can be accessed on shiny dashboard https://qakhuf-victor0pacifique-rwandarwacu.shinyapps.io/Shiny_survey_analysis/

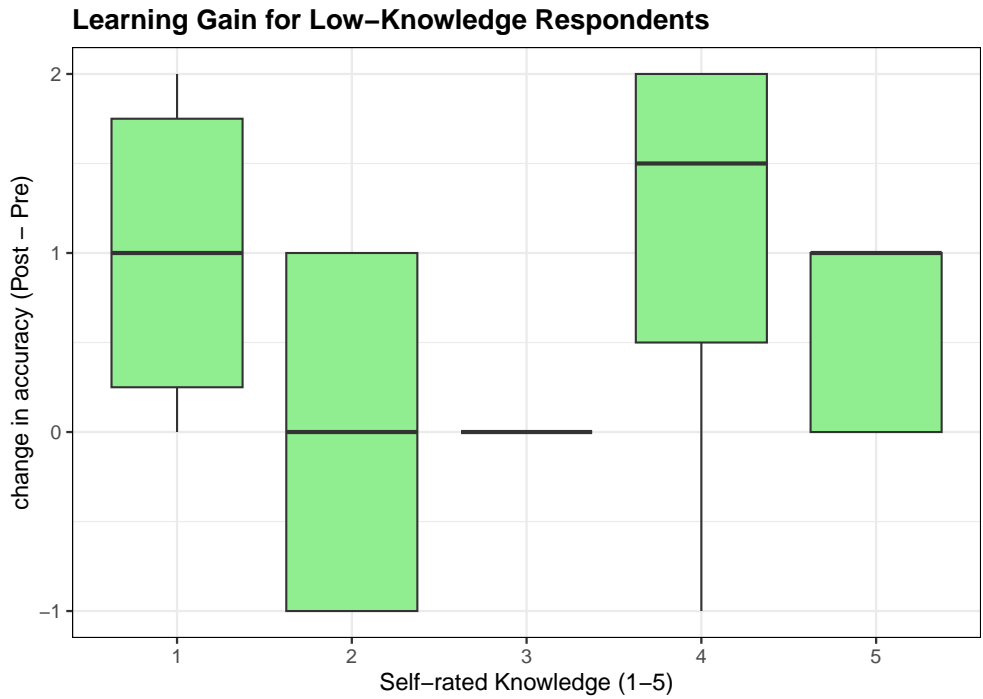

Figure 10: Learning Gain for Low-Knowledge Respondents: Overall there is remarkable gain post learning section

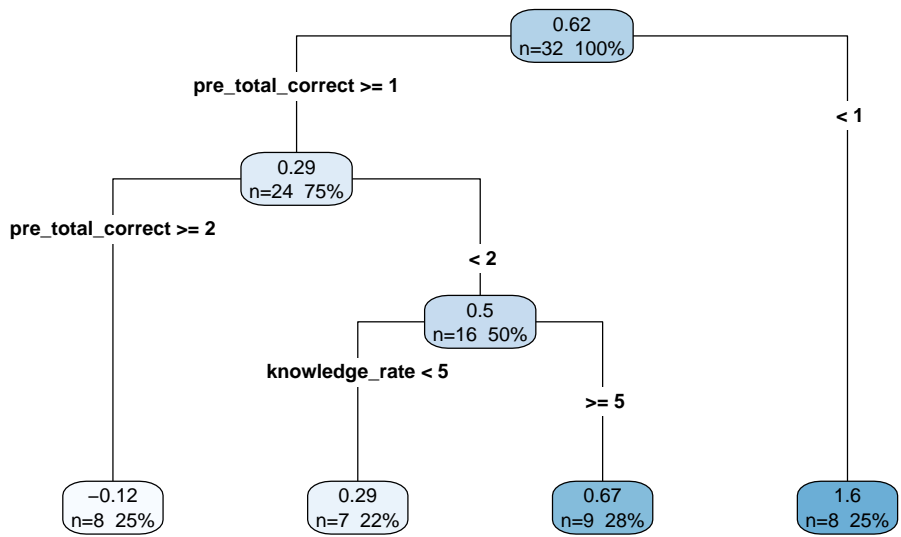

Figure 11: Regression Tree Predicting Learning Gain: respondent who score zero in pretest , had the highest learning gain

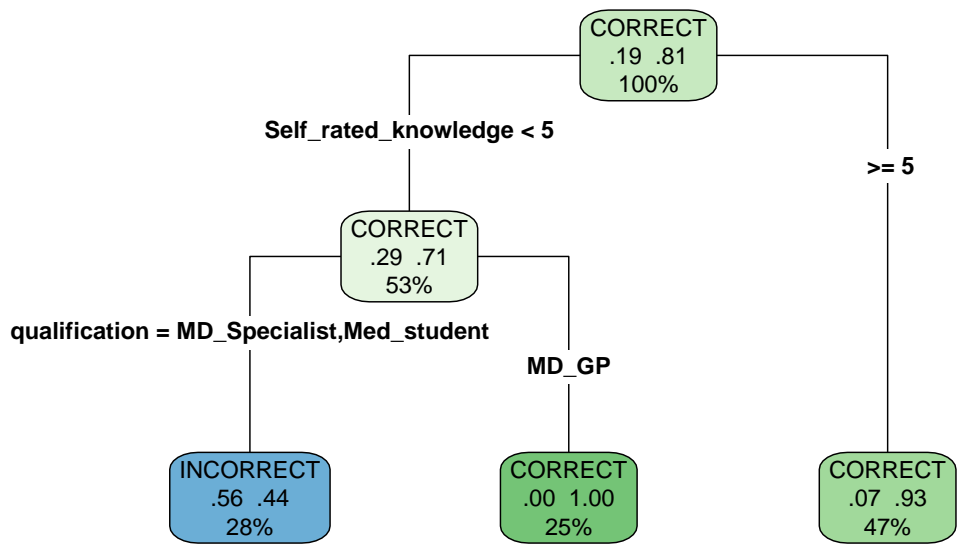

Figure 12: Prediction of correct MRL plot interpretation post learning section

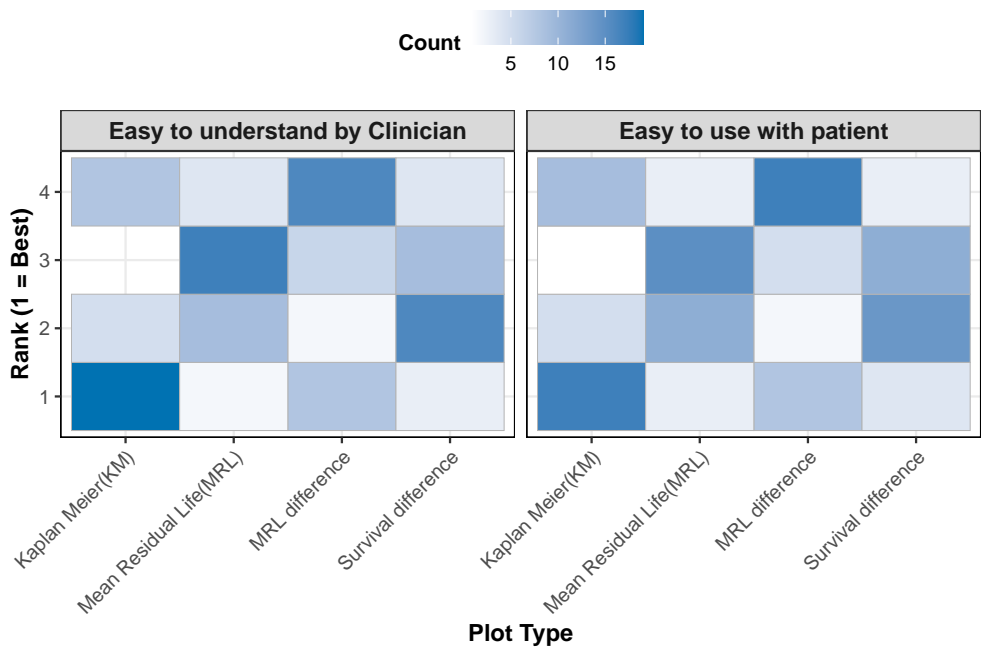

Figure 13: Ranking distribution

# 4 Discussion

## 4.1 Principal Findings

This pilot study assessed clinicians' and medical students' knowledge of survival visualisation, provided a learning section, evaluated learning outcomes, and ranked their preferences regarding the visualisations of survival and Mean residual life function. The study revealed that context-specific and straightforward guidance significantly improved participants' ability to interpret survival visualisation, particularly with the novel MRL plots, which were less familiar but ultimately more intuitive to many health professionals.

At baseline, interpretation accuracy was generally moderate, with some extremely low performance of less than 25%. The performance was lower for the novel Mean residual life plots. Following a review of correct interpretation, interpretation accuracy improved across all visualisations, with the most significant gains seen for the MRL plot, which showed an increase in correct responses from 43.8% to 81.2%. In comparison, KM plots increased from 56.2% to 81.2%. These findings highlight that, despite their initial complexity, MRL-based visualisations can become accessible and interpretable with minimal instruction. This supports previous claims that intuitive, time-based survival measures—such as MRL—can bridge the gap between statistical abstraction and clinical usability (Spiegelhalter et al. [2011]; Trevena et al. [2013]).

Furthermore, the MRL plot is the top alternative visualisation of KM for patient communication, despite being a novel function to the majority of participants. This implies that clinicians valued the "remaining time" framing for its resonance with patients' everyday reasoning. As expected, KM plots remained the most preferred visualisation for clinical use, likely due to their frequent use in medical research and their familiarity with health professionals through education on various levels, including those with limited statistical training.

The study's findings contribute to evidence-based medical practice and are consistent with a growing body of literature that advocates for more patient-centered approaches to presenting survival data. Kaplan–Meier plots, while standard in clinical trials and epidemiological research rely on cumulative probabilities over time, which are complex concepts for both clinicians and patients (Collett [2023]; Hawkins et al. [2005]). MRL plots offer a more concrete alternative by directly presenting the expected remaining time of survival, conditional on having survived to a specific point in time. This intuitive time-based metric aligns with what patients often seek to understand in the consultation: "How much longer do I have?" (Spiegelhalter et al. [2011]).

This work extends to the previous theoretical and methodological work by Alvarez-Iglesias et al. [2015] and Jalali [2018] on the hybrid estimator in the case of right-censored observations, which is commonly encountered in the health sector. The current study highlighted the ability and preferences of clinicians regarding both functions despite their limited exposure to the mathematical and statistical concepts; this study enhances the idea of translational statistics by involving end users' opinions on the choice of what facilitates their daily communication with patients, especially on counseling and patient education on clinical outcome.

## 4.2 Implications for Clinical Communication

This study reveals that integrating MRL visualisations into clinical communication tools could improve the clarity and relevance of doctor-patient discussions on clinical outcomes. For patient-centered care and informed decision-making, healthcare professionals require communication tools that combine statistical accuracy with accessibility. The MRL plot, with its direct expression of life expectancy at different time points, has the potential to enhance shared decision-making, particularly in chronic conditions or palliative care (Bélanger et al. [2011]; Wang et al. [2017]).

Moreover, the results indicate that even brief, clearly annotated visuals can significantly enhance understanding of complex survival concepts. Thus, including such visualization in curricula, especially those targeting evidence-based medicine and communication skills, may help bridge the gap between statistical training and practical application, as well as the dissemination of new evidence for practice. It is particularly encouraging that participants with lower self-rated knowledge of survival analysis achieved significant learning gains (Figure 9), reinforcing the potential for scalability in training.

The clear distinction between plots preferred for clinician use (KM) and those for patient discussion (MRL) highlights an opportunity for dual-mode visualisation strategies. Future clinical dashboards may include both types simultaneously, allowing clinicians to find a safety net that guarantees statistical precision and communicative clarity, depending on the audience and context.

### 4.3 Strengths and limitations of the study

This study offers multiple strengths to the literature. It presents the first empirical comparison of Kaplan-Meier and mean residual life visualization methods within a real-world clinical population, utilizing both pre-post testing and user preference rankings. Secondly, by examining both recurring errors and the interpretative approach, the mixed methods provided qualitative and quantitative perspectives on survival plot interpretation, yielding insights that could inform robust recommendations for the content and methodology of education on survival visualization designed explicitly for clinicians. The application of a semi-parametric hybrid MRL estimator guarantees that the visualizations are founded on robust and clinically relevant estimates, even when right-censored data is present.

Nonetheless, constraints are present. The limited sample size, resulting from convenience sampling, diminishes the generalizability of the findings. The sample inadequately represented the diversity of clinical specializations, age demographics, or educational backgrounds. The survey was administered online in a self-directed mode, which restricts control over participant participation and may result in discrepancies in attention or comprehension.

### 4.4 Recommendations

To inform clinical practice with robust evidence, this pilot study should be extended to a diverse group of healthcare providers, including not only medical doctors but also interdisciplinary teams such as nurses, allied health professionals, and even patients, to assess the effectiveness of the MRL plot compared to the conventional Kaplan-Meier (KM) plot, in a controlled environment. By including more participants, the sample size will increase, helping to determine the practical applicability and accessibility of MRL plots in real-world healthcare settings.

Moreover, given the existing demand for digital migration in the health system and telemedicine, integrating these visualisations into more interactive formats, such as clinical dashboards, could improve usability. By ensuring that all healthcare professionals have intuitive access to clear survival data, such tools may enhance interpretation and support more informed clinical responses. As visual analytics become increasingly embedded in electronic health records and decision support systems (Rostamzadeh et al. [2021]), MRL plots have the potential to play a central role in shaping the next generation of survival data communication.

## 5 Conclusion

This pilot study demonstrates that clinicians and medical students can significantly enhance their understanding of advanced survival visualizations, particularly Mean Residual Life (MRL) plots, by using contextual annotated graphs. Although Kaplan–Meier (KM) plots remain the most familiar and preferred for clinician use, MRL plots have emerged as the easiest to learn and are likely to be applied to patients. These findings highlight the need to explore different approaches for visualising health data, as well as the engagement of healthcare practitioners to understand which visualisations translate better to both clinicians and patients.

The significant learning gains observed for MRL plots highlight their potential as an educational and communication tool in clinical settings. Importantly, participants with lower self-rated knowledge of survival analysis achieved significant improvements, suggesting that MRL visualisation literacy can be readily scaled across diverse user groups regardless of prior statistical knowledge. However, a large-scale study conducted in a controlled environment is needed to generate robust evidence.

Overall, this study supports the integration of MRL visualisations into clinical practice and medical education, contributing to more transparent, intuitive, and patient-centered survival outcomes.

# 6 Acknowledgements

I would like to sincerely acknowledge the invaluable support and guidance of my supervisor, Prof. John Newell, whose expertise and encouragement were instrumental throughout this project. Appreciation is also due to Ranganath Shyamsundar and Niloofar Parastoo for their conceptual contributions and good collaboration. I also extend my gratitude to the staff and lecturers of the School of Mathematical and Statistical Sciences at the University of Galway for providing an exceptional and professional learning environment, and to all clinicians and medical students who participated in the survey and shared their time and insights.

# 7 Conflict of interest

There is no conflict of interest.